\documentclass[prd, preprint, nofootinbib]{revtex4}
\usepackage[dvipdfmx]{graphicx}%%
\usepackage{amsfonts}
\usepackage{latexsym}
\usepackage{amssymb}
\usepackage{epsfig}
\usepackage{comment}%\begin{comment}--\end{comment}
\usepackage{here}	
\usepackage{amsmath}

\def\slashb#1{\not\!\!#1}
\newcommand{\im}[1]{\text{Im}\,#1}

\begin{document}

\title{Modular forms of finite modular subgroups \\ from magnetized D-brane models}

\author{Tatsuo Kobayashi, and Shio Tamba}
 \affiliation{
Department of Physics, Hokkaido University, Sapporo 060-0810, Japan}

%\author{}
 %\affiliation{}
%\author{}
 %\affiliation{}

%\date{\today}

\begin{abstract}
We study modular transformation of holomorphic Yukawa couplings in 
magnetized D-brane models.
It is found that their products are modular forms, which are non-trivial representations of 
finite modular subgroups, e.g. 
$S_3$, $S_4$, $\Delta(96)$ and $\Delta(384)$.
\end{abstract}

\pacs{}
\preprint{EPHOU-18-014}
\preprint{}

\vspace*{3cm}
\maketitle

%====================================================%
%====================================================%
%   <<<<<<<<<<<<<<<<<<Main body>>>>>>>>>>>>>>>>>>>>      %
%====================================================%
%====================================================%

%==================================%
%                 1. Introduction               %
%==================================%

\section{Introduction}

The origin of the flavor structure in the quark and lepton sectors is one of unsolved but important mysteries in 
particle physics.
The quark and lepton masses have a hierarchy.
The lepton mixing angles are large, while the quark mixing angles are small.
Many studies have been done in order to understand this flavor structure.
Among them, non-Abelian discrete flavor symmetries are one of interesting approaches.
Indeed, many types of model building have been studied by use of various non-Abelian discrete groups such as 
$S_N$, $A_N$, $D_N$, $\Delta (3N^2)$, $\Delta(6N^2)$, etc.
(See for review \cite{Altarelli:2010gt,Ishimori:2010au,King:2013eh}.)

Superstring theory is a promising candidate for unified theory of all the interactions including 
gravity and matter particles such as quarks and leptons as well as the Higgs particle.
The four-dimensional (4D) low-energy effective field theory  of superstring theory has several symmetries.
In particular, some non-Abelian discrete flavor symmetries appear in 
the 4D effective field theory from superstring theory with certain compactification.
That is, heterotic string theory on orbifolds leads to 
non-Abelian discrete flavor symmetries, e.g.  $D_4$, and $\Delta (54)$ \cite{Kobayashi:2006wq}.
(See also Refs.~\cite{Kobayashi:2004ya,Ko:2007dz,Olguin-Trejo:2018wpw,Nilles:2018wex}.)\footnote{
In Ref.\cite{Beye:2014nxa}, a relation between gauge symmetries and non-Abelian flavor symmetries is discussed 
at the enhancement point.}
In addition, magnetized D-brane models as well as intersecting D-brane models in type II superstring theory can realize 
similar discrete flavor symmetries \cite{Abe:2009vi,Abe:2009uz,BerasaluceGonzalez:2012vb,Marchesano:2013ega,Abe:2014nla}.

On the other hand, torus and orbifold compactifications have the so-called modular symmetry.
Recently, in Ref.~\cite{Kobayashi:2017dyu,Kobayashi:2018rad}, modular transformation behavior of 
zero-modes was studied in magnetized D-brane models.(See also Ref.~\cite{Cremades:2004wa}.)
Such behavior was also studied in heterotic orbifold models \cite{Lauer:1989ax,Lerche:1989cs,Ferrara:1989qb}.
These modular transformations act non-trivially on zero-modes.
In this sense, the modular symmetry is a sort of flavor symmetries.
However, the modular symmetry also transforms Yukawa couplings as well as higher order couplings, 
while these couplings are trivial singlets under the conventional flavor symmetries.

One of interesting aspects is that the modular group $\Gamma$ includes finite subgroups such as $S_3$, $A_4$, $S_4$ and $A_5$, 
which have been used for flavor model building as mentioned above.
(See e.g. Ref.~\cite{deAdelhartToorop:2011re}.)
Inspired by these aspects, recently a new approach to the lepton mass matrices was proposed in Ref.~\cite{Feruglio:2017spp}.
The three generations of leptons are assigned to non-trivial representations of the finite modular subgroup $A_4$.
The couplings as well as neutrino masses are also assigned to modular forms, which are non-trivial representations under $A_4$.
Such an idea was extended in Refs.~\cite{Kobayashi:2018vbk,Penedo:2018nmg,Criado:2018thu,Kobayashi:2018scp,Novichkov:2018ovf} 
by use of other finite modular subgroups $S_3$ and $S_4$ in addition to $A_4$.

In these studies, the key ingredients are  the modular forms of weight 2, which are non-trivial 
representations of finite modular subgroups.\footnote{See for modular forms, e.g. Refs.~\cite{Gunning:1962,Schoeneberg:1974,Koblitz:1984}.}
Such modular forms are found for $S_3$ doublet, $A_4$ triplet, and $S_4$ doublet and triplet 
in Refs.~\cite{Feruglio:2017spp,Kobayashi:2018vbk,Penedo:2018nmg}.
In this paper, we study Yukawa couplings in magnetized D-brane models.
In particular, we study modular transformation of holomorphic Yukawa couplings, which are 
holomorphic functions of the modulus.
Using them, we construct modular forms of modular weight 2, which are non-trivial representations 
of finite modular subgroups.

This paper is organized as follows.
In section \ref{sec:magne-D}, we review the modular symmetry in magnetized D-brane models and study 
modular symmetry of holomorphic Yukawa couplings.
In section \ref{sec:modular-form}, we study products of holomorphic Yukawa couplings.
We show that they correspond to modular forms of finite modular subgroups.
Section \ref{sec:conclusion} is conclusion and discussion.

\section{Modular transformation in magnetized D-brane models}
\label{sec:magne-D}

In this section, we briefly review  modular transformation of zero-mode wavefunctions in 
magnetized D-brane models.
See for detail Refs.\cite{Cremades:2004wa,Kobayashi:2017dyu,Kobayashi:2018rad}.
Then, we study modular transformation of holomorphic Yukawa couplings.

\subsection{Zero-mode wavefunction}
\label{sec:zero-mode}

Here,  we give a brief review on zero-mode wavefunctions on the two-dimensional torus $T^2$ with magnetic flux \cite{Cremades:2004wa}.
For simplicity, we concentrate on $T^2$ with $U(1)$ magnetic flux.
We denote real coordinates by $x$ and $y$, 
and we also use the complex coordinate $z= x+ \tau y$, where 
$\tau$ is the complex modular parameter.
The metric on $T^2$ is given by 
\begin{equation}
g_{\alpha \beta} = \left(
\begin{array}{cc}
g_{zz} & g_{z \bar{z}} \\
g_{\bar{z} z} & g_{\bar{z} \bar{z}}
\end{array}
\right) = (2\pi R)^2 \left(
\begin{array}{cc}
0 & \frac{1}{2} \\
\frac{1}{2} & 0
\end{array}
\right) .
\end{equation}
In order to construct $T^2$, we identify $z \sim z +1$ and $z \sim z + \tau$.

We introduce the $U(1)$ magnetic flux $F$ on $T^2$, 
\begin{equation}
F = i\frac{\pi M}{\im{\tau}}  (dz \wedge d\bar{z}), 
\end{equation}
which corresponds to the vector potential,
\begin{equation}
A(z) = \frac{\pi M}{\im{\tau}} \im{(\bar{z}dz)}.
\end{equation}
Here we concentrate on vanishing Wilson lines.

We study the spinor field on $T^2$, which has  two components,  
\begin{equation}
\Psi(z,\bar{z}) = \left(
\begin{array}{c}
\psi_+ \\
\psi_- 
\end{array}
\right).
\end{equation}
On the above magnetic flux background, we solve the zero-mode equation 
\begin{equation}
i \slashb{D} \Psi = 0,
\end{equation} 
for the spinor field with the $U(1)$ charge $q$.
The magnetic flux should be quantized by the Dirac condition such that 
$qM$ is integer.
Either $\psi_+$ or $\psi_-$ has zero-modes exclusively for $qM \neq 0$.
If $qM$ is positive, only $\psi_+$ has $qM$ zero-modes, while $\psi_-$ has no zero-mode.
That is a chiral theory.
Their zero-mode profiles are given by 
\begin{equation}
\psi^{j,qM}(z) = \mathcal{N} e^{i\pi  qM z \frac{\im{z}}{\im{\tau}}} \cdot \vartheta \left[
\begin{array}{c}
\frac{j}{qM} \\
0
\end{array}
\right] \left( qM z, qM\tau \right),
\end{equation}
with $j=0,1,\cdots, (qM-1)$, 
where $\vartheta$ denotes the Jacobi theta function, 
\begin{equation}
\vartheta \left[
\begin{array}{c}
a \\
b
\end{array}
\right] (\nu, \tau) = \sum_{l \in {\bf Z}} e^{\pi i (a+l)^2 \tau} e^{2 \pi i (a+l)(\nu+b)} .
\end{equation}
The normalization factor $\mathcal{N}$ is given by 
\begin{equation}
\label{eq:normalization}
\mathcal{N} = \left( \frac{2\im{\tau} qM}{\mathcal{A}^2} \right)^{1/4}, 
\end{equation}
with $\mathcal{A}= 4 \pi^2 R^2 \im{\tau}$.

The ground states of scalar fields also have the same profiles as $\psi^{j,qM}$.
We can compute the Yukawa coupling including one scalar and two spinor fields, whose wavefunctions are written by 
 $\psi^{i,M}$, $\psi^{j,N}$, and $(\psi^{k,M'})^{*}$, by carrying out overlap integral of these wavefuctions.  
For simplicity, we have normalized the charges such as $q$ in a proper way and replace $qM$ by $M$. 
Note that the gauge invariance requires $M+N+M'=0$.
That is, $M'$ must be negative if both $M$ and $N$ are positive.
That is the reason why we consider  the wavefunction $(\psi^{k,M'})^{*}$ \cite{Cremades:2004wa}.
Then, 
their Yukawa couplings are given by the wavefunction overlap integral \cite{Cremades:2004wa},
\begin{eqnarray}
Y_{ijk} &=& g \int d^2 z \psi^{i,M}\psi^{j,N}(\psi^{k,M'})^{*}  \nonumber \\
 &=& g \left(\frac{2\im{\tau} }{ \mathcal{A}^2}\right)^{1/4} \sum_{m \in Z_{M'}} \delta_{k,i+j+Mm}  
\cdot \vartheta \left[
\begin{array}{c}
\frac{Ni-Mj+MNm}{MNM'} \\
0
\end{array}
\right] \left( 0, MNM'\tau \right),
\end{eqnarray}
where $g$ is constant.
Similarly, we can compute higher order couplings \cite{Abe:2009dr}.

\subsection{Modular transformation}

Here, we study modular transformation.
First we briefly review   modular transformation of zero-modes \cite{Kobayashi:2017dyu,Kobayashi:2018rad}.
(See also \cite{Cremades:2004wa}.)

The torus $T^2$ is constructed by ${\mathbb R}^2/\Lambda$, where 
the lattice $\Lambda$ is spanned by the vectors $(\alpha_1,\alpha_2)$, 
i.e., $\alpha_1 = 2\pi R$ and $\alpha_2 = 2 \pi R \tau$.
The same lattice is obtained by the following change of basis, 
\begin{equation}
\label{eq:SL2Z}
\left(
\begin{array}{c}
\alpha'_2 \\ \alpha'_1
\end{array}
  \right) =\left(
  \begin{array}{cc}
a & b \\
c & d   
\end{array}
\right) \left(
  \begin{array}{c}
\alpha_2 \\ \alpha_1
\end{array}
  \right) ,
\end{equation}
where $a,b,c,d$ are integer with satisfying $ad-bc = 1$.
That is $SL(2,Z)$ transformation.

The modular parameter $\tau = \alpha_2/\alpha_1$ transforms as 
\begin{equation}
\tau \longrightarrow  \gamma \tau = \frac{a\tau + b}{c \tau + d},
\end{equation}
under (\ref{eq:SL2Z}).
This transformation includes two important generators, $S$ and $T$,
\begin{eqnarray}
& &S:\tau \longrightarrow -\frac{1}{\tau}, \\
& &T:\tau \longrightarrow \tau + 1.
\end{eqnarray}
They satisfy the following relations,
\begin{equation}
\label{eq:S-ST}
S^2=1, \qquad  (ST)^3=1.
\end{equation}

If we impose further algebraic relation in addition to Eq.~(\ref{eq:S-ST}), 
we can realize finite modular subgroups.
For example, when we impose 
\begin{equation}
T^N=1,
\end{equation}
the modular subgroups are isomorphic to $S_3$, $A_4$, $S_4$ and $A_5$ for $N=2,3,4$, and 5, respectively.
Also, we can obtain other finite modular subgroups by imposing further algebraic relations.

Following \cite{Kobayashi:2017dyu,Kobayashi:2018rad}, we restrict ourselves to even magnetic fluxes $M$ ($M>0$).
Under $S$, the zero-mode wavefunctions transform as \cite{Cremades:2004wa,Kobayashi:2017dyu,Kobayashi:2018rad}
\begin{equation}
\label{eq:magne-S}
\psi^{j,M} \rightarrow \frac{1}{\sqrt{M}}\sum_k e^{2\pi i jk/M} \psi^{k,M}.
\end{equation}
On the other hand, the zero-mode wavefunctions transform as \cite{Kobayashi:2017dyu,Kobayashi:2018rad}
\begin{equation}
\label{eq:magne-T}
\psi^{j,M} \rightarrow e^{ \pi i j^2/M} \psi^{j,M},
\end{equation}
under $T$.
Generically, the $T$-transformation satisfies 
\begin{equation}
T^{2M} = 1,
\end{equation}
on the zero-modes, $\psi^{j,M}$.
Furthermore, in Ref. \cite{Kobayashi:2017dyu} it is shown that 
\begin{equation}
(ST)^3 = e^{\pi i/4},
\end{equation}
on the zero-modes, $\psi^{j,M}$ for generic case.

The holomorphic part of Yukawa couplings is written by 
\begin{eqnarray}
& & X^{i,M}(\tau) = \vartheta \left[
\begin{array}{c}
\frac{i}{M} \\
0
\end{array}
\right] \left( 0, M\tau \right).
\end{eqnarray}
We study modular transformation of $X^{i,M}(\tau)$.
Hereafter, we often denote $X^{i,M}(\tau)$ by $X^{i,M}$.
It is straightforward to study the modular transformation behavior of $X^{i,M}$ 
by using Eqs.~(\ref{eq:magne-S}) and (\ref{eq:magne-T}).
That is, the holomorphic function $X^{i,M}$ transforms 
\begin{equation}
\label{eq:Y-S}
X^{j,M} \rightarrow \sqrt{\frac{-i\tau}{M}}\sum_k e^{2\pi i jk/M} X^{k,M},
\end{equation}
under $S$, and 
\begin{equation}
\label{eq:Y-T}
X^{j,M} \rightarrow e^{ \pi i j^2/M} X^{j,M},
\end{equation}
under $T$.
Thus, the holomorphic Yukawa couplings as well as physical Yukawa couplings $Y_{ijk}$ 
transform non-trivially under the modular group.
Note that the $T$ transformation is diagonal in this basis.

\section{Modular forms of finite modular subgroups}
\label{sec:modular-form}

Here, we study construction of modular forms of weight 2 for finite modular subgroups.
The modular forms of level $N$ and weight $w$ are holomorphic functions of modulus $\tau$, which have 
the following modular transformation behavior:
\begin{equation}
f_i(\gamma \tau) = (c\tau + d)^w\rho(\gamma)_{ij}f_j(\tau),
\end{equation}
where $\rho(\gamma)$ is a unitary representation of $\Gamma_N=\Gamma/\Gamma(N)$ with a 
principal congruence subgroup $\Gamma(N)$.
The weights $w$ are even and modular forms of weight 2 are important, 
because other modular forms are obtained by their products.

Such modular forms  of weight 2 are expected to be derived by productions of $X^{j,M}$ 
like $X^{i,M} X^{j,M} X^{k,M} X^{\ell,M}$ 
as seen from the behavior (\ref{eq:Y-S}).
In what follows, we study such products concretely.

\subsection{ $M=2$}

The $S$ and $T$ transformations are represented on $X^{j,2}$ with $j=0,1$  by

\begin{equation}
\left( \begin{array}{c}
 X^{0,2} \\ 
X^{1,2} 
\end{array} \right) \longrightarrow \sqrt{-\tau}\rho(S)\left( \begin{array}{c}
 X^{0,2} \\ 
X^{1,2} 
\end{array} \right), \qquad \rho(S) = {\sqrt \frac{i}{2}}\left(
\begin{array}{cc}
1  & 1 \\
1  &  -1
\end{array}
\right),
\end{equation} 
and 
\begin{equation}
\left( \begin{array}{c}
 X^{0,2} \\ 
X^{1,2} 
\end{array} \right) \longrightarrow \rho(T)\left( \begin{array}{c}
 X^{0,2} \\ 
X^{1,2} 
\end{array} \right), \qquad \rho(T) = \left(
\begin{array}{cc}
1  & 0 \\
0  &  i
\end{array}
\right).
\end{equation} 
They satisfy the following algebraic relations,
\begin{equation}
\label{eq:ST-M=2}
\rho(S)^2=i, \qquad \rho(T)^4=1, \qquad (\rho(S)\rho(T))^3=-1.
\end{equation}
Similarly, represenatations of $S$ and $T$ on generic $X^{j,M}$ satisfy 
\begin{equation}
\label{eq:ST-M=2}
\rho(S)^2=i, \qquad \rho(T)^{2M}=1, \qquad (\rho(S)\rho(T))^3=-1.
\end{equation}

Now, let us study quartic tensor products of $X^{j,M}$ with $j=0,1$, i.e.,
\begin{equation}
(X^{0,2})^4, \qquad (X^{0,2})^3X^{1,2}, \qquad (X^{0,2})^2(X^{1,2})^2, \qquad X^{0,2}(X^{1,2})^3, \qquad (X^{1,2})^4
\end{equation}
These are five-dimensional representation of $S$ and $T$, which satisfy $\rho(S^2)=1$, $\rho((ST)^3)=1$, and  $\rho(T^4)=1$.
This five-representation is a reducible representation, and we can decompose to two irreducible representations, a doublet and a triplet.
The doublet corresponds to 
\begin{eqnarray}
Z_1=(X^{0,2})^4+(X^{1,2})^4, \qquad Z_2=2\sqrt{3}(X^{0,2})^2(X^{1,2})^2.
\end{eqnarray}
Then, on $(Z_1,Z_2)^T$, $S$ and $T$ are represented by
\begin{equation}
\label{eq:S3-2}
\rho(S) = \frac12 \left(
\begin{array}{cc}
-1 & -\sqrt{3} \\
-\sqrt{3} & 1 \\
\end{array}\right), \qquad 
\rho(T)=\left(
\begin{array}{cc}
1 & 0 \\
0 & -1 \\
\end{array}\right).
\end{equation}
They satisfy $\rho(S^2)=1$, $\rho(T^2)=1$, and $\rho((ST)^3)=1$.
That is nothing but $S_3$.
Thus, $(Z_1,Z_2)$ are the modular forms corresponding to the $S_3$ doublet.

The other three elements are written by 
\begin{eqnarray}
Z_3=(X^{0,2})^4-(X^{1,2})^4, \qquad Z_4=2\sqrt{2}(X^{0,2})^3X^{1,2}, \qquad Z_5=2\sqrt{2}X^{0,2}(X^{1,2})^3.
\end{eqnarray}
On $(Z_3,Z_4,Z_5)^T$, $S$ and $T$ are represented by 
\begin{equation}\label{eq:S4-3}
\rho(S) = \frac12 \left(
\begin{array}{ccc}
0 & -\sqrt{2} & -\sqrt{2} \\
-\sqrt{2} & -1 & 1 \\
-\sqrt{2} & 1 & -1 \\
\end{array}\right), \qquad 
\rho(T)=\left(
\begin{array}{ccc}
1 & 0 & 0\\
0 & i & 0 \\
0 & 0 & -i \\
\end{array}\right).
\end{equation}
They satisfy $\rho(S^2)=1$, $\rho(T^4)=1$, and $\rho((ST)^3)=1$.
That is isomorphic to $S_4$.
Thus, $(Z_3,Z_4,Z_5)$ corresponds to the $S_4$ triplet.

\subsection{$M=4$}

Similarly, we can study the $M=4$ case.
Note that $(ST)^3$ is trivial transformation on $\tau$, but 
$(ST)^3$ transforms the lattice vectors $(\alpha_1,\alpha_2)$ to $(-\alpha_1,-\alpha_2)$.
The wavefunctions satisfy the following relation,
\begin{equation}
\psi^{j,M}(-z) = \psi^{M-j,M}(z).
\end{equation}
Thus, for $M>2$ it is convenient to use the following basis,
\begin{equation}
\psi^{j,M}_{\pm}= \frac{1}{\sqrt{2}}( \psi^{j,M}(z) \pm  \psi^{M-j,M}(z)),
\end{equation} 
except $j=0, M/2$ 
in order to represent $S$.
That is the $Z_2$ orbifold basis \cite{Abe:2008fi}.
The $\psi^{j,M}_+$ are $Z_2$ even modes, while  $\psi^{j,M}_-$ are $Z_2$ odd.
The $\psi^{0,M}$ is always $Z_2$ even, and 
the $\psi^{M/2,M}$  is also $Z_2$ even when $M$ is even.

Similarly, we use the same basis, i.e., 
\begin{equation}
X^{j,M}_{\pm}= \frac{1}{\sqrt{2}}( X^{j,M}(z) \pm  X^{M-j,M}(z)),
\end{equation} 
except $j=0, M/2$.

When $M=4$, only $X^{1,4}_-$ is $Z_2$ odd, and it is singlet under the modular symmetry.
Now, we study the other $Z_2$ even elements, $X^{0,4}$, $X^{1,4}_+$, and $X^{2,4}$.
On $(X^{0,4},X^{1,4}_+,X^{2,4})^T$, $S$ and $T$ are represented by 
\begin{eqnarray}
\rho(S)=\frac{\sqrt{i}}{2}\left(
\begin{array}{ccc}
1 & \sqrt{2} & 1 \\
\sqrt{2} & 0 & -\sqrt{2} \\
1 & -\sqrt{2} & 1\\
\end{array}\right), \qquad \rho(T)=\left(
\begin{array}{ccc}
1& 0 & 0 \\
0 & e^{\pi i/4} & 0 \\
0 & 0 & -1 \\
\end{array}
\right).
\end{eqnarray}

Here, let us study quartic tensor products, 
\begin{equation}
(X^{0,4})^\ell (X^{1,4}_+)^m (X^{2,4})^n,
\end{equation}
with $\ell + m + n=4$.
Totally, there are fifteen elements.
They provide us a reducible representation of $S$ and $T$.
Thus, we decompose them to irreducible representations.
The simplest one is the singlet $Z_1$, which is written by 
\begin{equation}
Z_1 = \frac{2\sqrt{2}}{\sqrt{3}}((X^{1,4}_+)^4 - 2((X^{0,4})^3X^{2,4}+X^{0,4}(X^{2,4})^3)).
\end{equation}
On this singlet, $S$ and $T$ are represented by 
\begin{equation}
\rho(S)=-1, \qquad \rho(T) = -1.
\end{equation}

The next simplest irreducible representation is the doublet, $(Z_2,Z_3)$, which are written by 
\begin{eqnarray}
Z_2 &=& (X^{0,4})^4+(X^{2,4})^4 +6 (X^{0,4})^2(X^{2,4})^2, \nonumber \\
Z_3 &=& \frac{4}{\sqrt{3}}( (X^{1,4}_+)^4 + (X^{0,4})^3X^{2,4}+X^{0,4}(X^{2,4})^3).
\end{eqnarray}
On $(Z_2,Z_3)^T$, $S$ and $T$ are represented by the same $\rho(S)$ and $\rho(T)$ as Eq.~(\ref{eq:S3-2}).
Thus, this is the $S_3$ doublet.
Similarly, there is the triplet $(Z_4,Z_5.Z_6)$, which are written by 
\begin{eqnarray}
Z_4 &=& (X^{0,4})^4+(X^{2,4})^4 -2 (X^{0,4})^2(X^{2,4})^2, \nonumber \\
Z_5 &=& 2\sqrt{2}(X^{1,4}_+)^2 ((X^{0,4})^2 + (X^{2,4})^2), \\
Z_6 &=& 4\sqrt{2} X^{0,4}(X^{1,4}_+)^2 X^{2,4}. \nonumber
\end{eqnarray}
On $(Z_4,Z_5,Z_6)^T$, $S$ and $T$ are represented by the same $\rho(S)$ and $\rho(T)$ as Eq.~(\ref{eq:S4-3}).
Thus, this is the $S_4$ triplet.

At this stage, nine elements remain.
Among nine elements, the  $(Z_7,Z_8,Z_9)$ is a triplet, where 
\begin{eqnarray}
Z_7 &=& (X^{1,4}_+)^2 ((X^{0,4})^2 - (X^{2,4})^2), \nonumber \\
Z_8 &=& X^{0,4}X^{1,4}_+ ((X^{0,4})^2 - (X^{2,4})^2), \\
Z_9 &=&  -X^{2,4}X^{1,4}_+ ((X^{0,4})^2 - (X^{2,4})^2). \nonumber
\end{eqnarray}
On this triplet $(Z_7,Z_8,Z_9)^T$, $S$ and $T$ are represented by 
\begin{eqnarray}
\rho(S)=-\frac12 \left(
\begin{array}{ccc}
0 & \sqrt{2} & \sqrt{2} \\
\sqrt{2} & 1 & -1 \\
\sqrt{2} & -1 & 1 \\
\end{array}\right), \qquad 
\rho(T)=\left(
\begin{array}{ccc}
i & 0 & 0 \\
0 & e^{\pi i /4} & 0 \\
0 & 0 & e^{-\pi i /4} \\
\end{array}\right).
\end{eqnarray}
The other 6 elements correspond to an irreducible representation, 
$(Z_{10},Z_{11},Z_{12},Z_{13},Z_{14},Z_{15})$, where 
\begin{eqnarray}
Z_{10} &=& (X^{0,4})^4 - (X^{2,4})^4, \nonumber \\
Z_{11} &=& \sqrt{2} X^{0,4}X^{1,4}_+ ((X^{0,4})^2 +3 (X^{2,4})^2), \nonumber \\
Z_{12} &=&  \sqrt{2}X^{2,4}X^{1,4}_+ (3(X^{0,4})^2 + (X^{2,4})^2). \nonumber \\
Z_{13} &=& 2X^{0,4}X^{2,4}((X^{0,4})^2 - (X^{2,4})^2),   \\ 
Z_{14} & = & 2\sqrt{2}X^{0,4}(X^{1,4}_+)^3, \nonumber \\
Z_{15} & = & 2\sqrt{2}X^{2,4}(X^{1,4}_+)^3. \nonumber 
\end{eqnarray}
On them, $S$ is represented by 
\begin{eqnarray}
\rho(S)=-\frac12\left(
\begin{array}{cccccc}
0 & 1 & 1 & 0 & 1 & 1 \\
1 & 0 & 0 & 1 & 1 & -1\\
1 & 0 & 0 & 1 & -1 & 1 \\
0 & 1 & 1 & 0 & -1 & -1 \\
1 & 1 & -1 & -1 & 0 & 0 \\
1 & -1 & 1 & -1 & 0 & 0 \\
\end{array}\right),
\end{eqnarray}
and $T$ is represented by 
\begin{equation}
\rho(T) = {\rm diag}(1,e^{\pi i/4},-e^{\pi i/4},-1,e^{3\pi i/4},-e^{3\pi i/4}).
 \end{equation}
Both representations on $(Z_7,Z_8,Z_9)$ and $(Z_{10},Z_{11},Z_{12},Z_{13},Z_{14},Z_{15})$ satisfy 
\begin{equation}
\rho(S^2)=1, \quad \rho((ST)^3)=1, \quad \rho(T^8)=1, \quad \rho((ST^{-1}ST)^3)=1.
\end{equation}
This is the $\Delta(96)$ algebra.
In particular, $(Z_7,Z_8,Z_9)$ correspond to the $\Delta(96)$ triplet.

\subsection{Larger $M$}

Similarly, we can study the case with larger $M$.
Quartic tensor products of  $X^{j,M}$ would be modular forms of finite modular subgroups. 
We have obtained modular forms corresponding to $S_3 \simeq \Delta(6)$, $S_4 \simeq \Delta(24)$, 
$\Delta(96)$.
These are the  $\Delta(6N^2)$ series.
Thus, we may obtain modular forms of $\Delta(6N^2)$ with larger $N$ for larger $M$.

For example, we consider the case with $M=8$.
Then, we use the $Z_2$ even orbifold basis, i.e. 
$(X^{0,8},X^{1,8}_+, X^{2,8}_+,X^{3,8}_+,X^{4,8})$.
On them, $S$ is represented by 
\begin{eqnarray}
\rho(S)= \sqrt{\frac{i}{8}}\left(
\begin{array}{ccccc}
1 &  \sqrt{2} &  \sqrt{2} &  \sqrt{2} & 1 \\
 \sqrt{2} &  \sqrt{2} & 0 & - \sqrt{2}  & - \sqrt{2} \\
  \sqrt{2} & 0 & -2 & 0 &  \sqrt{2} \\
   \sqrt{2} & - \sqrt{2}  & 0 &  \sqrt{2} & - \sqrt{2} \\
   1 & - \sqrt{2} &  \sqrt{2} & -\sqrt{2} & 1 \\
   \end{array}\right),
   \end{eqnarray}
 and $T$ is represented by 
\begin{equation}
\rho(T) = {\rm diag}(1,e^{\pi i/8},e^{\pi i/2},e^{9\pi i/8},1).
 \end{equation}
Then, we consider quartic tensor products.
For instance, there is a triplet, $(Z_1,Z_2,Z_3)$, where 
\begin{eqnarray}
Z_1 &=& 2(X^{0,8}(X^{1,8}_+)^2X^{2,8}_+ + (X^{1,8})^2X^{2,8}_+X^{4,8}
-X^{0,8}X^{2,8}_+(X^{3,8}_+)^2-X^{2,8}_+(X^{3,8}_+)^2X^{4,8}), \nonumber \\
Z_2 & =&  (X^{0,8})^3 X^{1,8}_+ - X^{1,8}_+( X^{4,8})^3 +  (X^{0,8})^2 X^{1,8}_+ X^{4,8}  \nonumber \\
& & - X^{0,8} X^{1,8}_+  (X^{4,8})^2 +2  X^{0,8} (X^{2,8}_+)^2 X^{3,8} -2  (X^{2,8}_+)^2 X^{3,8}_+  X^{4,8}, \\
Z_3 & =&  -((X^{0,8})^3 X^{3,8}_+ - X^{3,8}_+( X^{4,8})^3 +  (X^{0,8})^2 X^{3,8}_+ X^{4,8} \nonumber \\
& & - X^{0,8} X^{3,8}_+  (X^{4,8})^2 +2  X^{0,8} (X^{2,8}_+)^2 X^{1,8} -2  (X^{2,8}_+)^2 X^{1,8}_+  X^{4,8}).
\end{eqnarray}
On them, $S$ and $T$ are represented by 
\begin{eqnarray}
\rho(S)=\frac12 \left(
\begin{array}{ccc}
0 & \sqrt{2} & \sqrt{2} \\
\sqrt{2} & -1 & 1 \\
\sqrt{2} & 1 & -1 \\
\end{array}\right),\qquad 
\rho(T)=\left(
\begin{array}{ccc}
e^{6\pi i/8 } & 0 & 0\\
0& e^{9 \pi i /8} & 0 \\
0 & 0& e^{\pi i /8}  \\
\end{array}\right).
\end{eqnarray}
They correspond to the $\Delta(384)$ triplet.
Similarly, we can discuss the case with larger $M$.

%==================================%
%        4. Discussion and conclusion        %
%==================================%

\section{Conclusion}
\label{sec:conclusion}

We have studied modular transformation of holomorphic Yukawa couplings 
in magnetized D-brane models.
Their products correspond to non-trivial representations of finite modular subgroups.
Explicitly, we have constructed modular forms of weight 2 for the 
$S_3$ doublet, the $S_4$ triplet, the $\Delta(96)$ triplet and sextet, and the 
$\Delta(384)$ triplet.
Similarly, these products with larger $M$ would give us representations 
for $\Gamma_{2M}$.

We have set vanishing Wilson lines.
In our discussion, the $Z_2$ orbifold basis is important.
The orbifold with magnetic flux allows discrete Wilson lines \cite{Abe:2013bca}.
It would be interesting to extend our analysis to 
 magnetized D-brane models with discrete Wilson lines.

Our results would be useful to study model building for the quark and leptoon masses 
and mixing angles by extending analyses in Refs.~\cite{Feruglio:2017spp,Kobayashi:2018vbk,Penedo:2018nmg,Criado:2018thu,Kobayashi:2018scp,Novichkov:2018ovf}.

%==========================================%
%<<<<<<<<<< ACKNOWLEDGMENTS >>>>>>>>>>>%
%==========================================%

\section*{Acknowledgments}
T.~K. was is supported in part by MEXT KAKENHI Grant Number JP17H05395.

%
%==========================================%
%<<<<<<<<<<< bibliography >>>>>>>>>>>>>%
%==========================================%

%%%%%%%%%%%%%%%%%%%%%%%%%%%%%%%%%%%%%%%%%%%%

\end{document}